# A DYNAMICAL ANALYSIS of the POOR GALAXY CLUSTERS ABELL 2626 and ABELL 2440[1]

Joseph J. Mohr[2,3], Margaret J. Geller[3] & Gary Wegner[4]

## ABSTRACT

We use 189 new radial velocities and $R$ band CCD photometry to study the galaxy clusters Abell 2626 and Abell 2440. By combining these new optical constraints with archival X-ray images and gas temperature measurements, we investigate the dynamical nature of both systems. We derive masses, luminosity functions, and mass-to-light ratios.

The symmetric X-ray emission from A2626 suggests a relaxed, single-component system; however, our sample of 159 redshifts reveals a complex, three component cluster at $cz \sim 17,500$ km/s. One of these components is a typical X-ray bright cluster, a second has a cluster-like galaxy population with a much lower central galaxy and gas density, and the third is a background structure with field galaxy composition. A comparison of the magnitude distributions within the two subclusters suggests that A2626 is a merger in progress (at 93% confidence). Virial masses and a composite luminosity function for a region with projected radius $r = 1.5h^{-1}$ Mpc yield a mass to light ratio $M/L_R \sim 610h$. Analysis of the X-ray emission from the primary component yields a gas mass fraction of $\sim 2.2h^{-3/2}\%$ and a baryon fraction of $\sim 3.4\%$. A radial infall model indicates that the virial mass may be an underestimate.

The bimodal X-ray emission and elongated galaxy distribution of A2440 have been studied before (Baier 1979; Beers *et al.* 1991). With deeper, CCD $R$ band photometry we demonstrate a striking correspondence between the galaxy and cluster gas distributions. The galaxy distribution has three main components, each associated with a giant elliptical galaxy. The two larger peaks in the galaxy distribution coincide with the primary peaks in the X-ray emission, and the third is associated with a significant X-ray surface brightness enhancement. We use this galaxy-gas correspondence, 48 redshifts, and the structure of the X-ray emission to argue that the subclusters are bound within a single system, and that the two primary components are beginning to merge. The composite luminosity function and estimates of the subcluster virial masses indicate a mass-to-light ratio in the range $M/L_R = 660h$ to $880h$.

# 1. INTRODUCTION

The strongest constraints on the dynamics of present epoch clusters come from combined optical and X–ray observations of the cluster galaxies and gas (e.g. Fabricant *et al.* 1986, 1989, 1993; Beers *et al.* 1991, 1992; Hughes 1989; Mohr *et al.* 1995; Burns *et al.* 1995; Zabludoff & Zaritsky 1995); these dynamical constraints provide critical tests for structure formation models (e.g. Richstone, Loeb & Turner 1992; Lacey & Cole 1993). It is important to quantify the evolution of poorer clusters along with the richer, typically better studied systems, because standard structure formation models predict differences in evolutionary history as a function of system mass (and perhaps by extension, richness) (e.g. Lacey & Cole 1995).

Here we report the results of a dynamical study of two poor clusters for which there are archival X–ray images and published gas temperatures. We supplement the available X–ray data with an extensive set of optical redshifts and mosaics of $R$ band CCD images. We use these data to study the structure of the two clusters. Their X–ray morphologies differ; A2626 has azimuthally symmetric emission centered on the dominant cluster elliptical, and A2440 has two bright X–ray peaks, each centered on a giant elliptical. The galaxy distributions reveal greater underlying complexity in both clusters. In A2626 a subcluster is evident in the projected galaxy distribution not as a separate peak, but as a significant asymmetry. In A2440 three galaxy condensations, each with an associated giant elliptical, are evident. In A2626, a redshift survey reveals a subcluster with significantly different mean velocity and velocity dispersion than the main component; in A2440 redshifts indicate that the subclusters have comparable mean velocities and are probably bound.

Along with improved constraints on substructure, a combination of optical and X–ray observations allows comparison of independent mass estimates and extraction of luminosity functions, mass–to–light ratios, gas masses and baryon fractions. This exercise is complicated by substructure; we consider each subcluster separately where possible. Both these poor clusters have apparently "high" mass–to–light ratios. The X–ray bright component of A2626 has a typical gas mass to galaxy mass ratio, but a "low" baryon fraction. We analyse the clusters as follows: section 2 focuses on A2626, section 3 on A2440 and section 4 contains a discussion of the primary results. Throughout the paper we use $H_0 = 100h$ km/s/Mpc.

# 2. ABELL 2626

Abell 2626 is a richness class 0 cluster (Abell 1958) with symmetric X–ray emission. Here we describe new optical spectroscopy and photometry which indicates that A2626 is composed of (a minimum of) two systems (§2.2). We use the X–ray emission and galaxy characteristics to consider the nature of the subsystems(§2.3); we calculate masses for the two main subcondensations, a composite luminosity function, mass–to–light ratios and the baryon fraction (§2.4). We then explore a merger hypothesis (§2.5).

## 2.1. Data

The appendix contains velocities and $R$ band photometry for 159 galaxies measured during the fall of 1993 and 1994 using the Decaspec (Fabricant & Hertz 1990) and the MkIII spectrograph mounted on the MDM 2.4m. The galaxies all lie within a projected distance of $1.5h^{-1}$ Mpc from the Abell cluster center. The spectra have 12Å resolution with coverage from 4500Å to 8500Å. The high signal–to–noise spectra yield a median (mean) velocity uncertainty of 45 km/s (50 km/s). This uncertainty includes the statistical uncertainty from the cross correlation with template spectra added in quadrature with the dispersion solution uncertainty determined from the positions of 4 sky lines in the complete sample of sky spectra. We add an additional uncertainty of 60 km/s in quadrature to velocities measured from emission lines; the emission line regions do not necessarily trace the galaxy center of mass (Thorstensen 1993; Kurtz *et al.* 1995). ZCAT (Huchra *et al.* 1992) contains 4 galaxy velocities within this region and we have remeasured velocities for at least three of them; the agreement is good. The fourth galaxy has a very poor ZCAT position, but its velocity and position are consistent with the large central elliptical in A2626.

We obtained $R$ band CCD photometry of the region within $1.5h^{-1}$ Mpc of the cluster center (diameter $\sim 1.5°$) with the Mt. Hopkins 1.2m in September 1994. This mosaic of 5 min exposures is non–photometric; we use a series of photometric 1.5 min calibration images (one short photometric image for every set of four deep images) and observations of Landolt (1992) standards to reduce the galaxy photometry to the Johnson–Kron–Cousins system. We acquired photometric images with the MDM 1.3m in October 1994. The $R$ band zero–point for each deep image is obtained using stars which appear in both the deep and the photometric overlap images; we then apply FOCAS software (Jarvis & Tyson 1981; Valdes 1982) to find galaxies and to determine their isophotal $R_{23.5}$ magnitudes. Each galaxy identified by FOCAS down to $R_{23.5} = 19$ is individually inspected to cull close double stars which tend to be misclassified as galaxies. The magnitude uncertainties are the statistical uncertainties in the isophotal magnitudes added in quadrature to a 0.035 mag uncertainty. This 0.035 mag uncertainty accounts for the flat fielding errors, the scatter in the $R$ band transformation between the deep and the photometric images, the scatter in the original $R$ band photometric solution (0.012 mag), and the MDM $R$ band filter color correction (we evaluate the color term $0.011(B-R)$ mag at $B-R=1$).

We evaluate the accuracy of the FOCAS magnitudes in two ways. First, we compare FOCAS magnitudes of galaxies which appear on more than one image; results indicate that the magnitude uncertainties are appropriate. Second, we compare FOCAS magnitudes with detailed, isophotal photometry carried out on deep, 2.4m images of 25 of the brighter early type galaxies (Mohr & Wegner 1996). One galaxy near a bright star has a FOCAS magnitude which is 0.5 mag too faint because of inaccurate FOCAS sky determination. The mean magnitude differences for the remaining 24 galaxies is -0.003 mag, and the scatter (0.079 mag) is roughly consistent with the quoted uncertainties.

The X–ray image of A2626 is a 2,050 s archival *Einstein* IPC observation. The reduced 0.3–3.5 keV image contains $\sim 620$ cluster photons. The image is reduced in the standard fashion (e.g. Mohr, Fabricant & Geller 1993) to a final resolution of $2.4'$ FWHM. David *et* *al.* (1993) list a cluster gas temperature of 2.9 keV ($1.5 \leq T \leq 11.0$ keV at 90% confidence) and a 2–10 keV luminosity of $1.91 \times 10^{43} h^{-2}$ ergs/s. Assuming a Raymond thermal spectrum (Raymond & Smith 1977; David *et al.* 1993) from a gas with 30% solar abundances (and including galactic absorption), we calculate a 0.5–3.5 keV luminosity of $3.71 \times 10^{43} h^{-2}$ ergs/s (using the PROS X–ray spectral package developed at SAO).

### 2.2. Partitioning the Cluster

The distribution of 159 velocities reveals a complex structure at $v \sim$17,500 km/s (see Figure 1). We partition the foreground structure into three systems. The gap in the velocity distribution at $\sim$20,500 km/s serves as a natural boundary. The galaxies between 14,000 km/s and 20,500 km/s appear to constitute two large systems with peaks at $\sim$16,500 km/s and $\sim$19,100 km/s. We examine the region (17,200 km/s$< v <$18,400 km/s) for evidence of an additional, smaller system. The eleven galaxies within this velocity range have spatial and velocity distributions consistent with their being outliers of the two main systems; additional galaxy radial velocities would further clarify this issue.

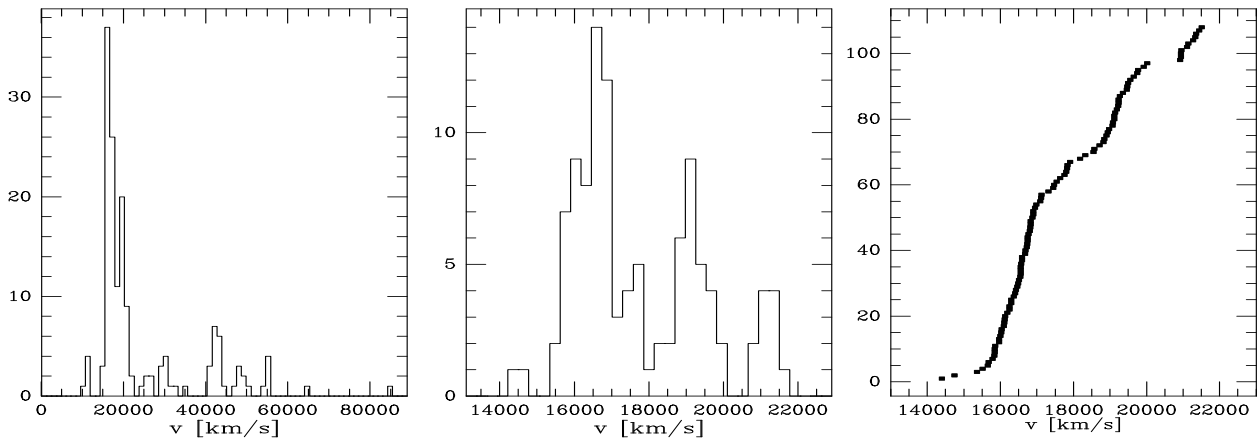

**Figure 1:** The galaxy velocities measured in A2626. From left to right we have (a) the histogram of all 159 measured velocities, (b) the histogram of the 108 near cluster velocities, and (c) the cumulative distribution of the near cluster velocities. The central, giant elliptical in group A has a velocity of 16,562±60 km/s.

We fit the velocity distribution between 14,000 km/s and 20,500 km/s to single and double Gaussians by minimizing the squared differences between the cumulative distributions, thereby avoiding bias associated with binning the observations. A KS test strongly rules out consistency between the observed velocity distribution and the best fit single Gaussian (consistency at a level of $2.3 \times 10^{-5}$ and $D_{KS} = 0.242$). The best fit double Gaussian model is an excellent description of the observations ($D_{KS} = 0.037$). We use this best fit double Gaussian model to partition the velocity data; the velocity at which there is a $\sim$50% chance of belonging to either system is 18,000 km/s. On the basis of the velocity distribution alone, a reasonable partition into three systems is (see Table 1): sixty–seven galaxies in the A group ($14,000$ km/s $< cz < 18,000$ km/s), thirty in the B group ($18,000$ km/s $< cz < 20,500$ km/s) and 11 in the C group ($20,500$ km/s $< cz < 22,000$ km/s). It is important to recognize that although this partition is sensible, the boundaries are approximate, and there is probably overlap of the individual velocity distributions.

Figure 2 indicates that although groups A & B are clustered on the sky, group C is not. A 2–D KS test clearly distinguishes among the spatial distributions of groups A, B & C (Press *et al.* 1992). We examine groups A and B for evidence of further substructure using the $\Delta$ statistic (Dressler & Shectman 1988). Varying the number of near neighbors from 4 to 10, we use the $\Delta$ statistic and fail to find evidence of substructure above the $2\sigma$ level in either system. Interestingly, the $\Delta$ statistic fails to detect substructure when groups A and B are considered together; this failure is caused by spatial mixing of the two galaxy populations and the grossly enhanced global dispersion (A and B combined yield $\sigma = 1,300$ km/s) which is the normalization for the statistic. When all three groups are considered, the statistic indicates (at 95% confidence) substructure associated with the group C galaxies in the NE corner of Figure 2.

### 2.3. Nature of the Subclusters

The clumping of the galaxies in groups A & B on the sky supports the proposition that they are subclusters; the X–ray emission is coincident with group A (see Figure 3). The galaxies in group C are more diffusely distributed on the sky. In addition, the spectra of group C members are predominantly emission spectra (8/11=73%); emission spectra are much rarer in groups A (26/67=39%) and B (10/30=33%). These population differences also support the hypothesis that groups A & B are subclusters, and group C is a lower density structure with field galaxy composition (Mohr *et al.* 1996).

The mean velocities of groups A & C differ by 4,600 km/s. Because of this large velocity difference, groups A & C are probably not bound, and the velocity difference corresponds to a Hubble flow distance of $\sim 46h^{-1}$ Mpc. This distance is large enough that we disregard possible dynamical interactions between groups A & C. The velocity difference between groups A & B may reflect either Hubble flow or an infall velocity associated with an impending merger; learning which hypothesis is correct will also define the dynamical relationship between groups B & C. Specifically, if groups A & B are merging then group C is at a large distance from them; if group B is located in the space between groups A & C, then determining the effects of group C on B becomes more difficult. We return to the Hubble flow versus merger issue in §2.5.

Figure 3 contains a contour plot of the smoothed galaxy distribution and the X–ray emission from the cluster. Interestingly, the observed X–ray emission from A2626 is symmetric and coincident (to within $40''$) with the large elliptical at the center of group A. This central elliptical in group A has a velocity of 16,562±60 km/s and no emission lines; the lack of azimuthal symmetry in the galaxy halo indicates that a secondary nucleus lying to the NE of the primary nucleus is part of the central elliptical rather than a projection along the line of sight. The velocity and position of this elliptical are coincident with the minimum of the group A potential. Interestingly, the brightest member of group B is an edge on disk galaxy with strong emission lines, a clear warp in the disk, and a velocity of 18,903±39 km/s.

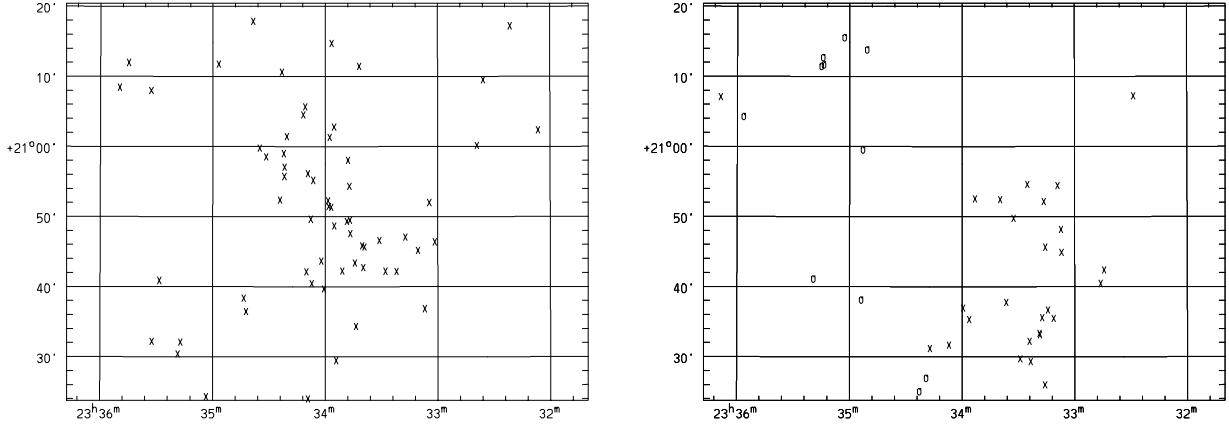

**Figure 2:** The positions of the 67 known members of clump A are marked on the left, and the positions of the 30 members of clump B (X's) and the 11 members of clump C (O's) are marked on the right. Group C has a high emission fraction (73%) typical of a field galaxy population; groups A and B are more clustered on the sky and have emission fractions ($\sim 35\%$) more typical of rich clusters.

Rather than being located $15'$ to the SW of the peak in the X–ray emission with the majority of group B members, this bright disk galaxy is just $1.3'$ west of the central elliptical in group A. Although many of the other galaxies in group B are early type, there is no "dominant" giant elliptical like the one in group A. To test whether group B is the edge of a larger cluster just outside our sample region we use the Digitized Sky Survey to examine the distribution of all galaxies brighter than $R = 16$ in a $3° \times 3°$ region centered on the Abell position; the bright galaxy distribution fails to support this hypothesis.

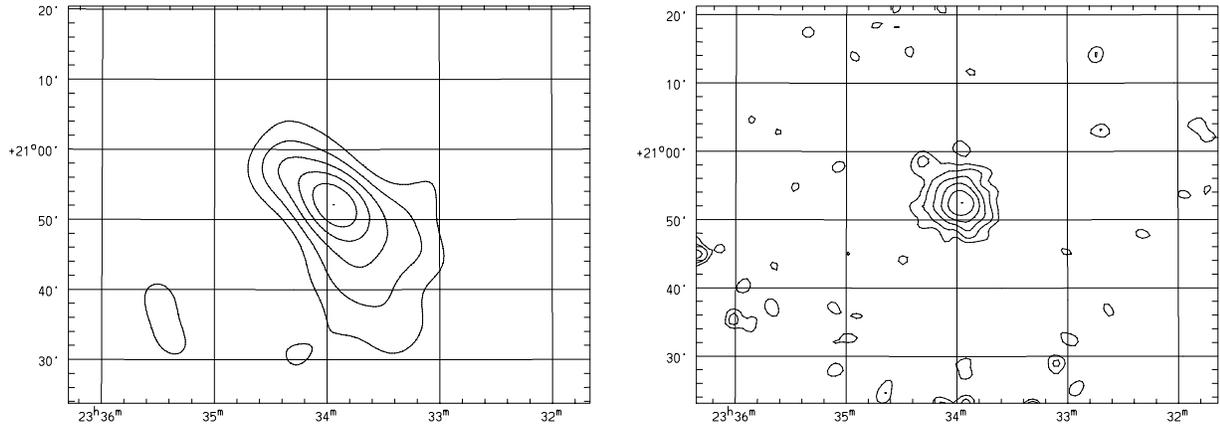

**Figure 3:** Left is a contour plot of the smoothed galaxy surface density distribution in A2626. All 169 unobserved galaxies with $R_{23.5} < 18$ and the 97 known group A and B members are included; the Gaussian smoothing length at a distance of 170 Mpc is $FWHM = 390$ kpc. The contours are linearly spaced between the peak of 1,218 gal/degree$^2$ and 428 gal/degree$^2$. The peak in the smoothed galaxy surface density lies within $40''$ of the position of the large, central elliptical; the extension to the SW of the peak is due to group B. Right is the contour plot of the *Einstein* IPC X–ray image of A2626. The contours are logarithmically spaced from a peak surface brightness of $1.5 \times 10^{-13}$ ergs/s/cm$^2$/arcmin$^2$ to a minimum surface brightness level of $7.9 \times 10^{-15}$ ergs/s/cm$^2$/arcmin$^2$. The resolution of this image at a distance of 170 Mpc is $FWHM = 118$ kpc.

The velocity dispersions of groups A & B are sensitive to the details of the partitioning. Using the partition described in §2.2, the dispersion for group A is $\sigma_A = 658^{+111}_{-81}$ km/s and the dispersion for group B is $\sigma_B = 415^{+117}_{-72}$ km/s (90% confidence). Ignoring the uncertainties associated with the partitioning, we note that an F–test indicates $\sigma_A > \sigma_B$ at 99.6% confidence.

The 2–10 keV X–ray luminosity of group A is $1.91 \times 10^{43} h^{-2}$ ergs/s (David *et al.* 1993), and there is no obvious X–ray emission from group B. The contrast in central surface brightness to the detector background determines the observability of an object. The *Einstein* IPC image of A2626 enables one to probe roughly 50× fainter than the peak of the emission from group A. Because the bremsstrahlung emissivity scales as the gas density squared and the emissivity within the IPC band is relatively insensitive to temperature variations (Fabricant, Lecar & Gorenstein 1980), this observability implies an upper limit to the central density of group B. The central gas density of group A is $\sim 3 \times 10^{-3} h^{1/2}$ cm$^{-3}$ (see §2.4); thus, the central gas density in group B must be $\leq 5 \times 10^{-4}$ cm$^{-3}$, just outside the range in X–ray bright clusters ($6 \times 10^{-4} h^{1/2}$ cm$^{-3}$ to $10^{-2} h^{1/2}$ cm$^{-3}$; Forman & Jones 1984). Although the contrast in the projected galaxy densities of groups A and B is difficult to quantify, the smoothed distribution (Figure 2) indicates that the central density in group B is $\sim 7$ times less than the central density in group A. Thus, the ratio of gas to galaxy density may be the same in both groups.

In summary, group C is consistent with a low density background structure. Groups A & B both appear to be subclusters with different physical characteristics. In particular, (1) group A is X–ray bright and group B is not, (2) the galaxy and gas distributions in group A are more dense than in group B, (3) group A has a larger velocity dispersion than group B, and (4) a giant elliptical galaxy coincides with the minimum in the group A potential; there is no giant elliptical cleanly associated with the center of group B.

Table 1: Abell 2626 Subsystems

| Group | N | $\bar{v}$ | $\sigma$ | $M[10^{14} M_\odot]$ | $M/L_R$ |
|---|---|---|---|---|---|
| A | 67 | 16,533±141 | $658^{+111}_{-81}$ | $6.6^{+2.4}_{-1.5}$ | 630 |
| B | 30 | 19,164±138 | $415^{+117}_{-72}$ | $2.3^{+1.4}_{-0.7}$ | 570 |
| C | 11 | 21,173±119 | $200^{+119}_{-52}$ | | |

Intervals are statistical 90% confidence limits

*2.4. Binding Masses, Gas Fraction and $M/L_R$*

The mass estimates (Table 1) assume distances proportional to the system mean velocity with a Hubble constant of 100 km/s/Mpc. The 90% confidence limits in the mean velocity, dispersions, and masses reflect purely statistical uncertainties (Danese, De Zotti & di Tullio 1980; Heisler, Tremaine & Bahcall 1985). The mass is determined using the mass estimators described by Heisler *et al.* (1985). These mass estimators can be systematically biased by the assumptions used to interpret the projected data. For reference we list the ratios of the virial, median, projected and mean masses (Table 1 contains the virial mass) which combine the projected positions and line of sight velocities in slightly different ways; the ratios for group A are 1:0.8:1.4:1.1 and for group B are 1:0.9:1.7:1.2. If the merger hypoethesis is correct and groups A and B are at a distance of 175 Mpc, the sum of their virial masses is $9.1^{+3.1}_{-2.7} \times 10^{14} M_\odot$ (90% confidence).

Assuming that the X–ray emitting gas in group A is in hydrostatic equilibrium, we use the X–ray emission to calculate a core mass (Bahcall & Sarazin, 1977; Fabricant et al. 1980). We fit the azimuthally averaged X–ray surface brightness profile to the standard $\beta$-model (Sarazin & Bahcall 1977, Cavaliere & Fusco–Femiano 1978). The parameters with a reduced $\chi^2 = 0.62$ fit to the 10 point radial profile (corresponds to 74% chance of consistency between the fit and the data) with $R \leq 0.529 h^{-1}$ Mpc are: $I_0 = 1.53 \times 10^{-13}$ ergs/s/cm$^2$/arcmin$^2$, $R_c = 0.143 h^{-1}$ Mpc, and $\beta = 0.843$. There is no spatially resolved temperature information in A2626, and the global temperature constraints are weak: $T = 2.9^{+8.1}_{-1.4}$ keV (90% confidence; David et al. 1993). The binding mass within the X–ray bright $0.5 h^{-1}$ Mpc cluster region is $M_X = 1.25 \times 10^{14} M_\odot$, and an extrapolated mass within $1.5 h^{-1}$ Mpc is $M_X = 4.0^{+11}_{-1.9} \times 10^{14} h^{-1} M_\odot$. Given this broad range, the hydrostatic, isothermal binding mass is consistent with the group A virial mass.

Using the bolometric X–ray luminosity (David et al. 1993) and the gas distribution, we determine the central gas density $n_0 = 3.5 \times 10^{-3} h^{1/2}$ cm$^{-3}$ (assuming 1/3 solar metallicity). The gas mass enclosed within $1.5 h^{-1}$ Mpc is $M_{gas} = 1.4 \times 10^{13} h^{-5/2} M_\odot$ and is uncertain by $\sim 25\%$. Both the central density and gas mass within $1.5 h^{-1}$ Mpc are consistent to within 25% of the values obtained in a revised Jones & Forman analysis of a large sample of clusters (Jones 1996). Using our measured gas mass, the corresponding gas mass fraction for group A is $\sim 2.2 h^{-3/2}\%$. This gas fraction is lower than the range ($3.5$–$7.8 h^{-3/2}\%$ within a radius of $0.5 h^{-1}$ Mpc) in 19 clusters observed with the IPC (White & Fabian 1995). Interestingly, eighteen of these nineteen clusters have higher Abell richness class than A2626.

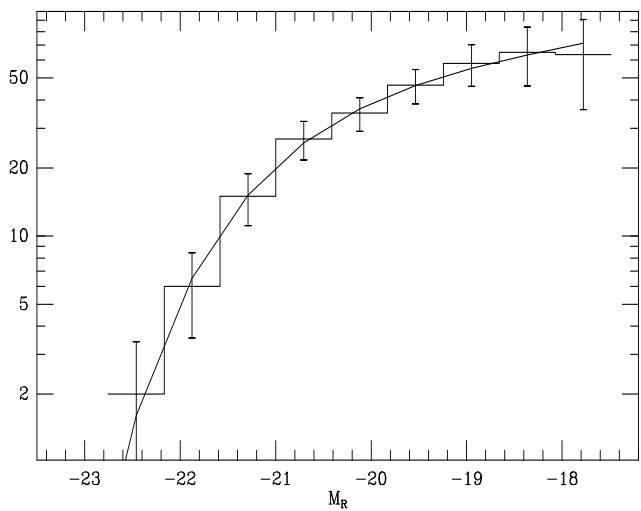

**Figure 4:** The composite luminosity distribution and best fit Schechter function for the A2626 galaxies brighter than $R_{23.5} = 19$. The best fit parameters are $M_R = -21.22^{+0.48}_{-0.44} + 5\log h$ and $\alpha = -1.16^{+0.18}_{-0.16}$ (90% statistical confidence limits). The associated total cluster $R$ band light is $L_R = 1.46^{+0.19}_{-0.23} \times 10^{12} h^{-2} L_\odot$, and the corresponding mass–to–light ratio when using the sum of virial masses (Table 1) is $M/L_R = 610 h$.

We approximate the total $R$ band light from the A2626 complex by fitting the galaxy magnitude distribution for $R_{23.5} < 19$ to a Schechter function. The Galactic extinction is $A_R = 0.0622$ (NASA/IPAC Extragalactic Database, Savage & Mathis 1979), and we use a uniform $k$ correction of $k = 0.0584$, appropriate for early type galaxies at a redshift $cz = 17,500$ km/s (Frei & Gunn 1994). Under the assumption that groups A and B are at a uniform distance of 17,500 km/s, the distance modulus to A2626 is $D = 36.34 - 5\log h$. Because we have a rather large redshift sample, we treat the faint end and the bright end differently. Specifically, for those galaxies with $R_{23.5} < 17$ we use only known cluster members and apply a completeness correction to each bin of the differential magnitude distribution. For the fainter galaxies ($17 < R_{23.5} < 19$), we include all galaxies imaged in our field and make a statistical background correction (Lopez-Cruz 1995). After the completeness correction and background subtraction the total sample of 766 galaxies brighter than $R_{23.5} = 19$ yields 318±34 "cluster" galaxies. The best fit Schechter parameters (Figure 4) are $M_R = -21.22^{+0.48}_{-0.44} + 5\log h$ and $\alpha = -1.16^{+0.18}_{-0.16}$. The total cluster $R$ band light is $L_R = 1.46^{+0.19}_{-0.23} \times 10^{12} h^{-2} L_\odot$ (Allen 1973). The uncertainties are 90% statistical confidence limits obtained by fitting 500 magnitude distributions randomly sampled from the best fit model. Combined with the sum of the virial masses (Table 1), the total cluster $R$ band mass–to–light ratio is $M/L_R = 610h$ in solar units. Using the 97% complete redshift sample brighter than $R_{23.5} = 16.25$, we determine that group A is emitting $\sim 72\%$ of the light and group B is emitting $\sim 28\%$. If these fractions remain roughly constant to fainter magnitudes the mass–to–light ratio for group A is $M/L_R = 630h$ and for group B is $M/L_R = 570h$.

Using a mass–to–light ratio typical for the cores of elliptical galaxies (Lauer 1985), we calculate the galaxy mass in group A, $M_{gal} = 8 \times 10^{12} h^{-1} M_\odot$. The uncertainties in this mass are dominated by the large variation in $M/L_R$ among ellipticals ($\sim 40\%$) and with galaxy type. Nevertheless, the ratio of gas to galaxy mass within $1.5h^{-1}$ Mpc is $\sim 1.8h^{-3/2}$, consistent with measurements in other clusters (David et al. 1990; Dell'Antonio, Geller & Fabricant 1995). Assuming that all the galaxy mass is baryonic, the baryon fraction for group A within $R < 1.5h^{-1}$ Mpc is $\sim 3.4\%$.

### 2.5. Merger Versus Hubble Flow

The 2,600 km/s velocity difference between the two groups could either be Hubble flow or a gravitationally induced peculiar velocity. There are three distinct configurations which could describe the two groups. Groups A & B could be (1) unbound– in which case the *minimum* distance between the two systems is the Hubble flow distance $l > 26h^{-1}$ Mpc, (2) bound with separation increasing– the separation $l > vt_H \sim 34$ Mpc, or (3) bound with separation decreasing– the distance to group B is less than (but comparable to) the distance to group A.

To determine which configuration best describes the two subclusters, we compare the apparent magnitude distributions (Lucey, Currie & Dickens 1986). This approach assumes that the luminosity distributions of the galaxies in groups A & B are intrinsically similar; a situation where the intrinsic differences in the cluster magnitude distributions perfectly *mask* a minimum 16% distance difference is obviously contrived. Thus, we regard this line of analysis as a means of *estimating* the probability that the two systems are at similar distances (gravitationally bound).

The velocity sample is 97% complete to $R_{23.5} = 16.25$, and it contains 45 galaxies from group A and 16 galaxies from group B. A KS test fails to distinguish between the two samples (see Figure 5); there is a 90% probability that the differences between the two distributions are consistent with statistical fluctuations ($D_{KS}$=0.16). A KS test also fails to distinguish between the total sample (48% complete to $R_{23.5} = 18$) of 67 galaxies in group A and 30 galaxies in group B.

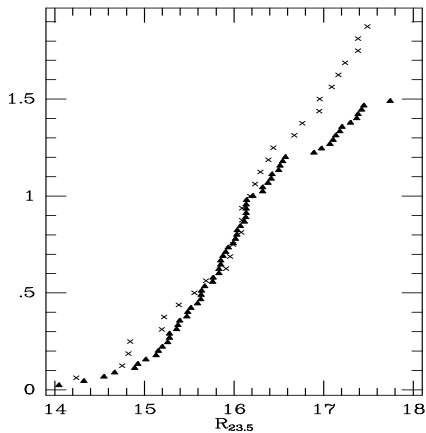

**Figure 5:** The cumulative magnitude distributions of the 67 galaxies in group A (triangles) and the 30 galaxies in group B ($x$'s). This galaxy sample is 97% complete to $R_{23.5} = 16.25$, and the cumulative distributions are normalized to the samples brighter than this limit. A KS test fails to distinguish between these two magnitude limited distributions. We extend the cumulative distribution beyond $R_{23.5} = 16.25$ to display the magnitude distributions of the incomplete portion of the sample.

To what extent are the similarities between the two magnitude distributions simply a result of small sample size (61 galaxies with velocities and magnitudes)? The mean velocities of groups A and B differ by 16%; if this velocity difference were simply Hubble flow then the difference in distance modulus would be $\sim 0.32$ mag. Assuming that the luminosity distributions of groups A & B are similar, we test whether our sample would reveal a 0.32 mag offset. First we simply shift the group B sample 0.32 mag brightward; the KS distance between the two samples (to $R_{23.5} = 15.9$) increases, indicating the Hubble flow hypothesis is less consistent with the data than the "merger" hypothesis. We evaluate the significance of this result by (1) choosing subsamples of 16 galaxies from the 32 galaxies in the complete group A sample with $R_{23.5} < 15.93$, (2) shifting each subsample 0.32 mag faintward, and then (3) looking at the distribution in $D_{KS}$ between the 45 galaxies in the group A sample with $R_{23.5} < 16.25$ and the 16 galaxies in the shifted sample. If the group A and B magnitude distributions were identical but offset by 0.32 mag, there would be a 93% chance (930 of 1,000 trials) of obtaining a larger $D_{KS}$ than observed. Thus, the magnitude distributions of the galaxies in groups A and B are sufficient to rule out the Hubble flow hypothesis at the 93% confidence level.

An examination of the radial infall model (Beers, Geller & Huchra 1982) indicates that there are no bound solutions given the virial masses of the individual systems and the line of sight velocity difference. If the subclusters are at the same distance, as indicated by the magnitude comparison, the radial infall model indicates that the virial masses of groups A and B are underestimates of the total binding mass (the sum of the virial masses must be larger by at least 65%); a confirmation of the same distance hypothesis requires knowledge of the true separation of the two clusters along the line of sight. Mohr & Wegner (1996) use fundamental plane distances to a sample of galaxies in groups A and B to measure this line of sight separation; results favor the merger hypothesis.

### 3. ABELL 2440

Abell 2440 is a richness class 0 cluster (Abell 1958) with clear bimodal structure in its X–ray emission. Baier (1979) studied the angular distribution of the cluster galaxies; Beers *et al.* (1991) use velocities of a sample of 24 cluster galaxies and the structure of the X–ray emission to study the cluster dynamics. They decompose their set of 24 galaxy velocities into three subsamples according to the projected distance of galaxies from three giant ellipticals and conclude that there is ample evidence for substructure. We use 30 additional galaxy velocities, deeper CCD photometry, and the morphology and temperature of the X–ray emitting gas to further constrain the cluster dynamics. We use the new data (§3.1) to probe for substructure (§3.2) and partition the sample; we compute the virial and hydrostatic mass estimates, the composite mass–to–light ratio, and discuss evidence for interactions between the two primary subclusters (§3.3).

*3.1. Data*

The appendix contains $R$ band photometry and 30 new galaxy velocities measured with the MMT red channel spectrograph in July 1994. The spectra have a resolution of $\sim 11$Å and coverage from 3800Å to 7400Å. Twenty–nine of these galaxies lie within the cluster. Three of the new velocities are for galaxies with previous measurements. The new measurements and the previous values are consistent within the uncertainties; we list the variance weighted composite velocity for these three galaxies. Two of the 24 galaxies with previously measured velocities have inaccurate positions. We discard those velocities rather than attempting to choose among the galaxies in the vicinity of the quoted position. Thus, the complete velocity sample consists of 48 cluster members and one (near) velocity outlier.

We include new galaxy photometry obtained from 15 minute $R$ band images of 8 contiguous fields within the central regions of A2440. The photometry comes from observations at the Mt. Hopkins 1.2m telescope in September 1994. The mosaic contains a $2 \times 2$ grid of $11'$ fields centered on the cluster and an additional 4 surrounding images. The skies were non–photometric during these observations. Therefore, we use four 1.5 minute images, taken in photometric conditions with the 1.2m telescope during May 1995, to determine a consistent photometric zero–point in the 8 deep fields. As in A2626 (see §2.1), we determine the zero–point and extract the isophotal galaxy photometry to $R_{23.5} = 20$. Four galaxies with measured velocities lie outside the region covered with CCD photometry; we use magnitudes determined from a $30' \times 30'$ scan of the POSS plates. We determine a mean offset between plate magnitudes and $R_{23.5}$ using 84 galaxies with measured plate and CCD photometry. The RMS scatter in this offset (0.24 mag) is reflected in the magnitude uncertainties.

The X–ray image we use for this analysis is a 2,822 s *Einstein* IPC archival image. The X–ray emission in the 0.3–3.5 keV band during the observation resulted in $\sim 900$ detected cluster photons. We reduce the image in the standard fashion (e.g. Mohr, Fabricant & Geller 1993). The final resolution after Gaussian smoothing is $2.4'$ FWHM. Figure 6 contains a contour plot of the X–ray emission with contours linearly spaced from a peak surface brightness of $2.3 \times 10^{-13}$ ergs/s/cm$^2$/arcmin$^2$ to a minimum of $1.1 \times 10^{-14}$ ergs/s/cm$^2$/arcmin$^2$ (corrected for cosmological dimming). In addition, Figure 6 contains a contour plot of the image signal to noise. From the cluster center out the contours correspond to $5\sigma$, $3\sigma$, and $1\sigma$, where the uncertainties include statistical noise and a 20% uncertainty in the level of the subtracted background. We note that there is also an archival ROSAT PSPC image of A2440, but the cluster lies well outside the central region where the PSF is small compared to that of the IPC; we do not include this image in the following analysis.

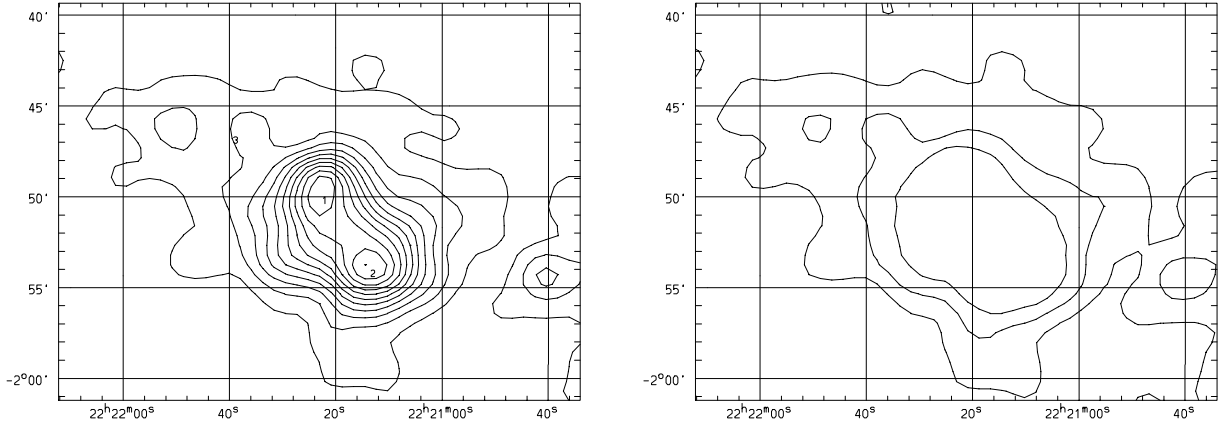

**Figure 6:** A contour plot (left) of the 0.3–3.5keV X–ray emission from Abell 2440 as measured with the *Einstein* IPC. The contours are linearly spaced from a peak surface brightness of $2.3 \times 10^{-13}$ ergs/s/cm$^2$/arcmin$^2$ to a minimum of $1.1 \times 10^{-14}$ ergs/s/cm$^2$/arcmin$^2$. Numbers mark the positions of three giant elliptical galaxies discussed in the text. The right contour plot shows the image signal to noise. From the image center out the contours correspond to $5\sigma$, $3\sigma$, and $1\sigma$.

From *Einstein* MPC data, David *et al.* (1993) measure an angularly unresolved X–ray temperature of 9.0 keV ($> 3.2$ keV at 90% confidence) and a 2–10 keV X–ray luminosity of $1.17 \times 10^{44} h^{-2}$ ergs/s. Assuming a Raymond thermal spectrum (Raymond & Smith 1977) with 30% solar abundances and the appropriate galactic absorption, we calculate a 0.3–3.5 keV X–ray luminosity of $1.18 \times 10^{44} h^{-2}$ ergs/s.

### 3.2. Evidence of Substructure

The most notable evidence for substructure in A2440 is its double peaked X–ray morphology. There are three giant, cluster ellipticals with small projected separation from the X–ray bright core: G1 ($\alpha$=22:21:22.46 $\delta$=-1:50:14.3, cz=26,972±25 km/s), G2 ($\alpha$=22:21:13.36 $\delta$=-1:54:14.5 cz=26,823±27 km/s) and G3 ($\alpha$=22:21:39.24 $\delta$=-1:46:56.2 cz=27,967±31 km/s). The positions of the ellipticals G1 and G2 are coincident with the peaks in the X–ray surface brightness (see Figure 6). There is also an X–ray surface brightness enhancement (at the $3\sigma$ level with central surface brightness $\sim 10$ times less than the main peaks) coincident with elliptical G3.

The smoothed galaxy distribution strengthens the case for cluster substructure; it reveals that each of the giant ellipticals is associated with a peak in the projected galaxy density. With our deeper photometry, the galaxy distribution around the two X–ray bright clumps changes from the single–peaked, elliptical structure first noted by Baier (1979) to well resolved, double peaks centered on the galaxies G1 and G2. The lower contrast peak associated with G3 is present as an extension of the main peak in the work by Baier. Figure 7 contains contour plots of the galaxy surface density distribution. Because only 48 galaxies have measured velocities we include all galaxies brighter than a magnitude cutoff; both contour plots are Gaussian smoothed to the same resolution as the X–ray image ($2.4' \sim 190$ kpc). Seventy–eight galaxies with $R_{23.5} < 18$ appear in the left contour plot. The contours are linearly spaced from a smoothed peak of 3,318 gal/degree$^2$ to 1,327 gal/degree$^2$. The right contour plot contains 169 galaxies with $R_{23.5} < 19$. The contours are linearly spaced from a smoothed peak of 6,160 gal/degree$^2$ to 2,464 gal/degree$^2$.

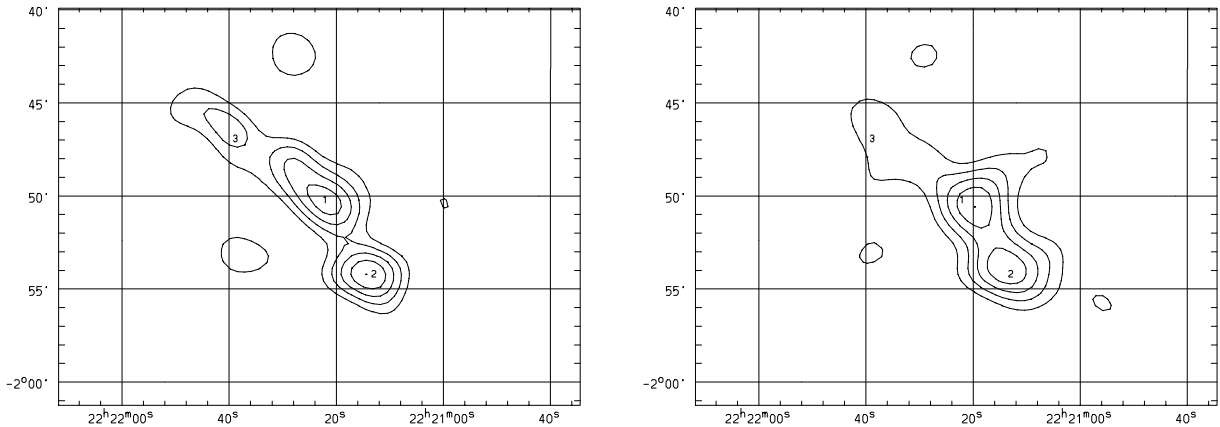

**Figure 7:** Contour plots of the Gaussian smoothed ($FWHM = 2.4' \sim 190$ kpc) galaxy surface density distribution to $R_{23.5} = 18$ (left) and $R_{23.5} = 19$ (right) in Abell 2440. The contours are linearly spaced from the peaks to 40% the peak value. The positions of the three giant ellipticals discussed in the text are marked.

The expected field galaxy density (Lopez–Cruz 1995) is 174 gal/degree$^2$ at $R \sim 18$ and 515 gal/degree$^2$ at $R \sim 19$. Thus, uncertainties in the galaxy distribution morphology are dominated by (1) variations in the galaxy populations as a function of the magnitude limit, (2) chance superpositions of background clusters and (3) sampling noise in the cluster galaxy distribution. We show two contour plots of the morphology to underscore the variations with depth. The differences are caused by an apparent deficit in the faint galaxy population in group G3 relative to groups G1 and G2. Similar variations in the faint end slope among groups have been noted elsewhere (Ferguson & Sandage 1991). We estimate the effects of sampling noise given the effective area of the Gaussian smoothing kernel, the galaxy surface density associated with the lowest contour in both plots, and an estimate of the mean galaxy density outside the contiguous peaks. Assuming the galaxy "background" is Poisson distributed about the mean, we calculate that there should be roughly 8 (3) spurious peaks in the left (right) contour plots at the level of the outer isodensity contour; only the three main peaks are statistically significant.

Figure 8 contains the velocity histogram and cumulative velocity distribution for the 48 cluster galaxies with measured velocities. The mean velocity for the entire sample is 26,997±253 km/s and the redshift corrected velocity dispersion (Danese *et al.* 1980) is $957^{+199}_{-136}$ km/s (90% statistical confidence limits). Galaxy G1 has a velocity consistent with the cluster center of mass, and galaxy G2 has a velocity roughly one $\sigma$ less than the cluster mean. G3 has a recession velocity that exceeds the cluster mean by 970 km/s. It appears that the velocity of G3 coincides with a small secondary peak in the velocity distribution; clearly, more galaxy velocities are required for confirmation. The best fit Gaussian distribution has a mean of 27,331 km/s and a dispersion of 1,305 km/s. A KS test disproves the hypothesis that the best fit Gaussian and the observed distribution come from the same parent distribution at the 93% confidence level.

Interestingly, the Dressler–Shectman statistic calculated using the 10 nearest neighbors (Dressler & Shectman 1988) fails to find any evidence for substructure in the sample of 48 galaxies. Out of 1000 Monte Carlo reshufflings of velocities, 881 produce a greater $\Delta_{10}$ than the observed distribution. However, when one uses the 4 nearest neighbors only 7 of 1000 reshufflings have higher $\Delta_4$'s. The signal comes from the galaxy G3 and its nearest neighbors; of the group of 5 galaxies two have newly measured velocities, and one is reobserved and confirmed. These new velocities, the third through fifth nearest neighbors of G3 strengthen the "low–mass clump" discussion by Beers *et al.* (1991). Given the obvious bimodality in the X–ray emitting gas, our failure to find substructure with the Dressler–Shectman test around the galaxies G1 and G2 underscores the restrictive nature of the test. A substructure detection with $\Delta$ requires spatially correlated velocity offsets or velocity dispersion offsets. Clearly, such a configuration does not describe all clusters undergoing mergers.

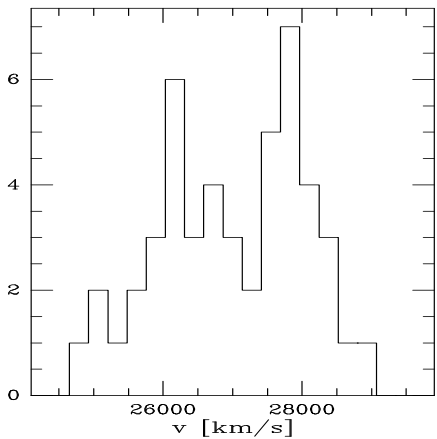

**Figure 8:** The distribution of 48 galaxy velocities in Abell 2440. The departure from a Gaussian is significant at 93% confidence.

In summary, the projected galaxy distribution, the X–ray emitting gas distribution, the locations and velocities of three giant ellipticals, and the line of sight velocity distribution all provide evidence for substructure in A2440. The similar nature of the projected galaxy and gas distributions is particularly striking. The cluster consists of a minimum of three locally bound subgroups.

### 3.3. Subcluster Masses, $M/L_R$, and Evidence for Interactions

Mass estimates relying on equilibrium assumptions are dangerous within systems with obvious substructure; nevertheless, the equilibrium assumption may be appropriate for each individual subcluster. Unfortunately, the projected separations of the three galaxy peaks on the sky are comparable to the scale lengths of the galaxy distribution associated with each; therefore, any partitioning by position on the sky mixes galaxies from the three main components. Furthermore, the velocity distribution does not suggest a clear partitioning of the sample. However, both the X–ray emitting gas and the galaxy distribution define three significant peaks, and each of these peaks is centered on one of the giant ellipticals near the cluster core. Thus, we investigate a cluster partition like that implemented by Beers *et al.* (1991).

We divide the velocity sample into three subsamples by associating each galaxy with the closest of the three giant ellipticals; group membership is noted for the galaxies listed in the Appendix. Table 2 lists the number of galaxies in the partition, the velocity of the associated giant elliptical, the mean and dispersion of each velocity distribution and the virial mass estimates (Heisler *et al.* 1985). The mass estimates assume a distance for all three subclumps of 270 Mpc, and the intervals listed are 90% statistical confidence limits. The uncertainties associated with the assumptions made in interpreting the projected positional and velocity data are difficult to quantify; we simply note the ratio of the virial, median, projected, and average mass estimators for each galaxy grouping: all 48 galaxies 1:1:1.5:1.2, group G1 1:1:1.5:1.2, group G2 1:1.3:2.1:1.8, and group G3 1:1.3:1.4:1.2. The virial mass for the combined sample of 48 galaxies is in good agreement with the results in Beers *et al.* (1991) for a sample of half the size. The only exception to the partitioning described above is in group G3 where we have excluded one very low velocity member which substantially affects the dispersion (galaxy number 15 in Beers *et al.* 1991). The contamination from incorrectly assigned membership tends to bias the mean velocities and dispersions toward similar values; although the results for the three different groups vary, these differences are not significant. The velocities of the giant ellipticals G1 and G2 are consistent with their group velocities; however, the velocity of G3 is $\sim 3\sigma$ away from its group mean. Interestingly, although G3 and its four nearest neighbors have very similar velocities, the nine other near neighbors have velocities more consistent with the global cluster distribution. Given the uncertainties in assigning group membership we interpret this offset as probable contamination from non–group G3 members.

Using the radial X–ray surface brightness profile of the clump centered on G2 within a wedge which excludes the emission associated with the G1 and G3 clumps, we estimate a radial fall–off and core radius ($\beta = 0.81$, $R_c = 0.12$ Mpc). The image resolution is taken into account in determining this core radius. Assuming hydrostatic equilibrium, we use these measurements and the global gas temperature ($T = 9.0$ keV, $> 3.2$ keV at 90% confidence; David *et al.* 1993) to constrain the radial mass distribution. Our approach yields a binding mass within 1 Mpc of the center of the G2 group of $8 \times 10^{14} M_\odot$, larger than the virial mass

Table 2: Abell 2440 Systems

| Partition | N | $v_{gal}$ | $\bar{v}$ | $\sigma$ | $M[10^{14}M_\odot]$ |
|---|---|---|---|---|---|
|  | 48 |  | 26,997±253 | $957^{+199}_{-136}$ | $6.3^{+2.9}_{-1.7}$ |
| G1 | 18 | 26,972±25 | 27,188±378 | $846^{+339}_{-181}$ | $2.9^{+2.8}_{-1.1}$ |
| G2 | 15 | 26,823±27 | 26,643±537 | $1,083^{+499}_{-250}$ | $3.3^{+3.7}_{-1.3}$ |
| G3 | 14 | 27,967±31 | 27,295±376 | $728^{+354}_{-173}$ | $2.2^{+2.5}_{-0.9}$ |

Intervals are statistical 90% confidence limits estimates listed for G1 and G2 in Table 2. However, the poorly constrained gas temperature allows for consistency among the mass estimates.

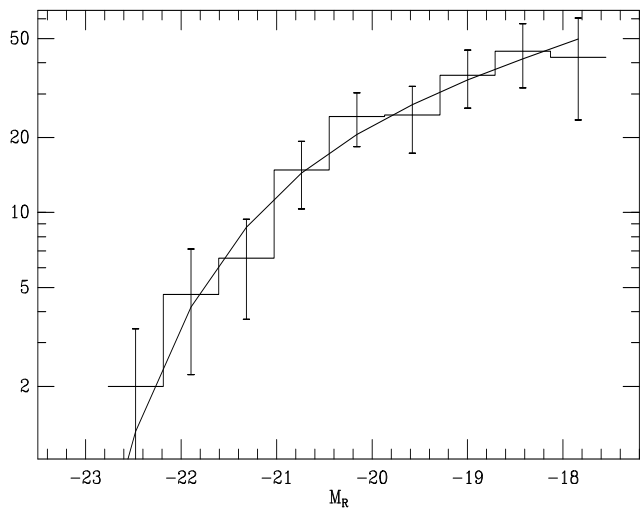

**Figure 9:** The composite luminosity distribution and best fit Schechter function for the background corrected galaxies brighter than $R_{23.5} = 20$ in Abell 2440. The best fit parameters are $M_R = -21.53^{+0.84}_{-0.64} + 5\log h$ and $\alpha = -1.30^{+0.26}_{-0.20}$ (90% statistical confidence limits). The associated total cluster $R$ band light is $L_R = 9.6^{+0.6}_{-1.5} \times 10^{11} h^{-2}\ L_\odot$, and the corresponding mass to light ratio when using the composite virial mass in Table 2 is $M/L_R = 660h$.

We calculate the cluster $R$ band light by fitting a Schechter function to the luminosity distribution within the whole cluster field. The Galactic extinction is $A_R = 0.085$ mag (NASA/IPAC Extragalactic Database, Savage & Mathis 1979), and we use a uniform $k$ correction of $k = 0.090$ mag, appropriate for early type galaxies (Frei & Gunn 1994). The distance modulus to A2440 is $D = 37.34 - 5\log h$. Using statistical background subtraction (Lopez-Cruz 1995) on a sample of 480 galaxies brighter than $R_{23.5} = 20$, we find 199±27 "cluster" galaxies with best fit Schechter parameters $M_R = -21.53^{+0.84}_{-0.64} + 5\log h$ and $\alpha = -1.30^{+0.26}_{-0.20}$. The total cluster $R$ band light is then $L_R = 9.6^{+0.6}_{-1.5} \times 10^{11} h^{-2}\ L_\odot$. The uncertainties in the best fit luminosity function parameters reflect the 90% confidence statistical uncertainties obtained through fitting 500 random samples of the best fit function. The limits on the total light reflect the 90% range in the luminosity function parameters constrained to give the correct normalization. Combined with the composite virial mass the cluster light yields a mass to light ratio $M/L_R = 660h$ in solar units. If we instead use the sum of the virial masses from G1, G2, and G3 we obtain $M/L_R = 880h$.

The existence of three distinct galaxy groups with comparable mean velocities implies that A2440 is dynamically young. In particular, cluster substructure should be erased over timescales comparable to the cluster crossing time $\sim 1$ Gyr (White & Rees 1978). However, similarity of the projected galaxy and gas distributions indicates that we are viewing this cluster before the collision of the three subclusters (Pearce, Thomas & Couchman 1994); the different dynamical behavior of the galaxies and the gas leads to short lived displacements during a merger (e.g. Zabludoff & Zaritsky 1995; Burns et al. 1995) which are not apparent in this cluster. The X–ray emission asymmetry which extends to the SE between groups G1 and G2 (significant at $\geq 5\sigma$) strongly suggests that there are ongoing gas interactions between groups G1 and G2. Although this emission excess could be evidence of another subcluster projected along the line of sight, there is no evidence for an associated galaxy clump.

## 4. DISCUSSION

We first summarize the evidence for substructure within these two poor clusters and then discuss the relatively high mass–to–light ratios and low baryon fraction. Interestingly, although standard models of cluster growth predict that substructure should be rarer in low mass clusters (e.g. Lacey & Cole 1995), we find strong evidence for continuing growth in both these poor clusters. Studies of representative cluster samples may provide surprising constraints on theories of structure formation.

### 4.1. Substructure Within Two Poor Clusters

With 159 new galaxy velocities and $R$ band photometry within $1.5h^{-1}$ Mpc of the cluster core, we show that A2626 is composed of two subclusters. Groups A and B have similar galaxy populations, but differ in other respects. Group A has a velocity dispersion of 658 km/s, is X–ray bright, and has a central, giant elliptical with two nuclei. Group B has a lower velocity dispersion of 415 km/s, lacks observable X–ray emission, and has no central elliptical. The densities of galaxies and gas in group B are much lower than those in group A, and the mean velocities of the two groups differ by $\sim 2{,}500$ km/s. A radial infall model combined with the virial masses of the two subclusters suggests that there is not enough cluster mass to bind groups A and B in a single system. However, a comparison of the galaxy magnitude distributions of groups A and B indicates that a merger is favored over Hubble flow at 93% confidence. The ultimate resolution of this contradiction lies in measuring the line of sight distance between the two subclusters (Mohr & Wegner 1996); if the merger hypothesis is correct then the virial mass must be a significant underestimate of the total binding mass.

The low density galaxy distribution in group B and the lack of X–ray emission may indicate that a first pass in the merger of the two subclusters has already taken place. The strongest argument against this model is that the X–ray emission is symmetric and well centered on group A. Interestingly, the brightest member of group B is an edge on, gas rich disk galaxy projected $1.3'$ from the dominant elliptical in group A (for comparison, the center of group B lies roughly $15'$ from the center of group A). Because gas rich galaxies with emission lines avoid the cores of clusters (e.g. Dressler 1980, Mohr et al. 1996), it may be that this galaxy is one of the first members of group B to pass through the core of group A. One might then expect a burst of star formation in this galaxy driven by the interactions between the cluster and galactic gas (e.g. Bothun & Dressler 1986). It is also possible that the bright disk galaxy is a velocity outlier of group A and has no physical association with group B, but the galaxy is located $3.6\sigma_A$ away from the mean velocity of group A.

We have used deep $R$ band photometry and 48 redshifts (30 new) in A2440 to develop a clearer picture of its dynamics. The cluster contains three subcondensations, each associated with a giant elliptical. The similar mean velocities and small projected separations of these subclusters strongly suggest they form a single, bound system. The correspondence between the galaxy and gas distributions is striking and suggests that we are viewing this cluster before significant collisional interactions among the subclusters. A feature in the X-ray emission extending to the SE from between the two main clumps probably indicates some interactions between the gas distributions of subclusters G1 and G2.

### 4.2. Masses and Luminosity Functions

We use the galaxies with measured velocities to calculate virial masses for the subclusters within both A2626 and A2440 (see Table 1 and Table 2) and compare these masses to hydrostatic, isothermal binding mass estimates where possible. The independent mass measures are consistent, but the uncertain global gas temperatures in both clusters weaken this statement. For each cluster we calculate a composite $R$ band luminosity function using all galaxies brighter than an absolute magnitude of $M_R \sim -17.5$. The best fit Schechter function parameters are $M_R = -21.22^{+0.48}_{-0.44} + 5\log h$ and $\alpha = -1.16^{+0.18}_{-0.16}$ for A2626 and $M_R = -21.53^{+0.84}_{-0.64} + 5\log h$ and $\alpha = -1.30^{+0.26}_{-0.20}$ for A2440 (90% confidence intervals). These values lie within the range defined by a study of 9 Abell clusters (Lugger 1986).

Using the luminosity functions and virial masses calculated over the same region of the cluster (a region with radius $1.5h^{-1}$ Mpc in A2626 and a smaller, more complicated region in A2440; see §3.1), we then estimate $R$ band mass–to–light ratios for both clusters. We find $M/L_R = 610h$ in A2626 and $M/L_R = 660h$ (or $880h$ depending on which virial mass is used) in A2440. In A2626, we use a magnitude limited sample of bright galaxies with redshifts to estimate the light contribution from each subcluster and find that the mass–to–light ratio of each subcluster is consistent with the composite value.

To review, we find good agreement between the $R_{23.5}$ magnitudes and more detailed, bulge plus disk decomposition photometry on a sample of 25 galaxies observed with the MDM 2.4m (see §2.1). We find that the mean difference between the isophotal $R_{23.5}$ and extrapolated total magnitudes is 0.24 mag (with a scatter of 0.14 mag) for the same galaxy sample. Thus, $R_{23.5}$ underestimates the total galaxy light by $\sim 25\%$. However, the total cluster light overestimates the light emitted from the virialized region probed by the mass estimators, because it includes light from emission line galaxies which exist outside the core. In the particularly well observed cluster A576, the emission line galaxies contribute 25% of the $R$ band light (Mohr *et al.* 1996). The total cluster light we calculate includes no correction for the emission line galaxies or for the isophotal magnitudes. If the mass–to–light ratios in A2626 and A2440 are representative of the mean mass–to–light ratio of the universe then $\Omega_0 \sim 0.4$–0.6 (Lin *et al.* 1996; they also use isophotal magnitudes).

These mass–to–light ratios exceed the median for groups and clusters within the CfA redshift survey (e.g. Ramella *et al.* 1989); however, approximately 25% of the groups and clusters in the CfA study have mass–to–light ratios higher than A2626 and A2440. Recent analyses of distant clusters using weak lensing techniques, which do not depend on the equilibrium assumption, yield mass–to–light ratios ranging from $300h$ (Tyson & Fischer 1995; Squires *et al.* 1996) to $\sim 700h$ (Fahlman *et al.* 1994; Carlberg, Yee, and Ellingson 1994; Luppino & Kaiser 1996). A comprehensive study of the masses (employing independent measures) and luminosity functions (relying on deep, CCD photometry) of a representative sample of clusters could provide interesting cosmological constraints.

With the X–ray bright subcluster in A2626 we use the details of the X–ray emission to calculate the gas mass and a mean galaxy $M/L_R$ (appropriate for ellipticals) to calculate the galaxy mass. Our gas mass is within 25% of that measured in an independent study of the same cluster (Jones 1996). The ratio of the gas to galaxy mass is $\sim 1.8h^{-3/2}$, typical of other clusters (David *et al.* 1990); however, the gas mass fraction is $2.2h^{-3/2}\%$, less than the fraction in a sample of 19 clusters studied with the *Einstein* IPC (White & Fabian 1995), but similar to the value for A576 (Mohr *et al.* 1996). Assuming all the galaxy mass is baryonic, the baryon fraction is $\sim 3.4\%$; if this baryon fraction is representative of the mean value in the universe, nucleosynthesis implies $\Omega_0 \sim 0.4$ (White *et al.* 1993).

## ACKNOWLEDGEMENTS

We thank Dan Fabricant for a critical reading of the manuscript that led to significant improvements, John Thorstensen for obtaining photometric overlap images in A2626, and Christine Jones for discussions and comparison of our gas mass with her unpublished results. We also express our appreciation to Bob Barr at the MDM Observatory, and John McAfee and Dennis Smith at the MMT Observatory for their assistance. Thanks also goes to Larry David for lending us his central gas density code. This work was supported in part by the NASA GSRP, NAGW–201, and NAGW–2367.

Appendix: Galaxy Velocities and Magnitudes

| RA (1950) | Decl | $R_{23.5}$ | $\sigma_R$ | $v$ | $\sigma_v$ | RA (1950) | Decl | $R_{23.5}$ | $\sigma_R$ | $v$ | $\sigma_v$ |
|---|---|---|---|---|---|---|---|---|---|---|---|
| | | | | | Abell 2626 | | | | | | |
| 23 33 59.48 | 20 52 09.2 | 14.04 | 0.036 | 16562 | 60 | 23 35 21.04 | 20 59 49.1 | 16.46 | 0.037 | 30287 | 38 |
| 23 33 54.10 | 20 52 25.9 | 14.24 | 0.036 | 18903 | 39 | 23 34 20.25 | 20 26 46.5 | 16.46 | 0.038 | 20956 | 42 |
| 23 35 18.21 | 20 31 56.7 | 14.32 | 0.036 | 17129 | 40 | 23 35 01.10 | 20 52 25.1 | 16.48 | 0.037 | 55282 | 42 |
| 23 34 29.30 | 20 22 22.3 | 14.54 | 0.036 | 17539 | 38 | 23 34 10.21 | 20 56 02.5 | 16.50 | 0.037 | 17442 | 36 |
| 23 34 22.71 | 20 56 55.6 | 14.66 | 0.036 | 16731 | 33 | 23 32 47.29 | 20 38 57.7 | 16.50 | 0.037 | 30016 | 36 |
| 23 34 26.20 | 20 57 26.8 | 14.70 | 0.036 | 11405 | 38 | 23 35 19.60 | 20 30 15.3 | 16.51 | 0.037 | 16028 | 37 |
| 23 33 08.31 | 20 48 02.8 | 14.75 | 0.036 | 19186 | 31 | 23 35 33.39 | 20 32 01.9 | 16.54 | 0.037 | 17117 | 31 |
| 23 33 57.29 | 20 35 11.2 | 14.82 | 0.036 | 19113 | 43 | 23 33 02.63 | 20 46 16.1 | 16.57 | 0.037 | 16983 | 44 |
| 23 33 16.90 | 20 25 50.1 | 14.84 | 0.100 | 18568 | 42 | 23 35 43.11 | 21 02 13.2 | 16.62 | 0.037 | 34941 | 40 |
| 23 34 08.75 | 20 49 29.9 | 14.89 | 0.036 | 17637 | 32 | 23 34 23.86 | 20 24 51.3 | 16.63 | 0.038 | 21102 | 37 |
| 23 34 24.15 | 21 10 32.3 | 14.92 | 0.036 | 15836 | 36 | 23 36 07.66 | 21 01 16.2 | 16.65 | 0.101 | 11891 | 70 |
| 23 32 40.07 | 21 00 01.6 | 15.01 | 0.036 | 16005 | 38 | 23 32 57.55 | 20 32 58.0 | 16.67 | 0.037 | 30424 | 35 |
| 23 33 48.15 | 20 54 12.0 | 15.13 | 0.036 | 16897 | 58 | 23 34 00.50 | 20 36 48.7 | 16.67 | 0.037 | 18847 | 55 |
| 23 33 49.52 | 20 49 12.1 | 15.15 | 0.036 | 16090 | 64 | 23 34 54.80 | 20 37 58.0 | 16.69 | 0.037 | 20936 | 70 |
| 23 34 07.96 | 20 31 29.6 | 15.19 | 0.036 | 19742 | 35 | 23 33 19.65 | 20 32 59.0 | 16.76 | 0.037 | 18944 | 38 |
| 23 35 29.48 | 20 40 44.5 | 15.20 | 0.036 | 16746 | 45 | 23 31 53.34 | 20 58 45.4 | 16.76 | 0.038 | 26459 | 44 |
| 23 33 19.80 | 20 33 12.0 | 15.22 | 0.036 | 19273 | 39 | 23 33 54.91 | 20 23 52.5 | 16.77 | 0.037 | 42960 | 70 |
| 23 35 14.90 | 21 11 34.8 | 15.23 | 0.036 | 21334 | 38 | 23 32 21.47 | 21 14 24.0 | 16.87 | 0.037 | 30106 | 37 |
| 23 36 38.38 | 20 46 19.4 | 15.26 | 0.036 | 16809 | 37 | 23 34 25.23 | 20 52 15.1 | 16.89 | 0.037 | 15364 | 40 |
| 23 34 32.46 | 20 18 14.5 | 15.27 | 0.037 | 17840 | 46 | 23 33 18.38 | 20 35 26.0 | 16.95 | 0.037 | 19236 | 42 |
| 23 33 40.11 | 20 45 31.4 | 15.28 | 0.036 | 17911 | 61 | 23 33 49.18 | 20 21 30.0 | 16.95 | 0.101 | 18322 | 47 |
| 23 35 33.91 | 21 07 54.7 | 15.35 | 0.036 | 16670 | 33 | 23 33 45.24 | 20 43 13.7 | 16.97 | 0.037 | 16556 | 37 |
| 23 32 46.13 | 21 16 37.8 | 15.37 | 0.037 | 11556 | 41 | 23 33 08.73 | 20 47 40.4 | 17.01 | 0.037 | 42372 | 45 |
| 23 33 47.81 | 20 49 21.8 | 15.37 | 0.036 | 16743 | 34 | 23 34 03.13 | 20 59 58.3 | 17.02 | 0.037 | 40969 | 37 |
| 23 33 17.68 | 20 22 54.3 | 15.38 | 0.100 | 19219 | 35 | 23 33 40.29 | 20 39 24.9 | 17.06 | 0.037 | 42080 | 46 |
| 23 33 23.09 | 20 42 02.1 | 15.39 | 0.036 | 15966 | 49 | 23 35 03.84 | 21 15 26.8 | 17.06 | 0.037 | 20947 | 34 |
| 23 33 22.38 | 20 18 49.5 | 15.47 | 0.100 | 17803 | 72 | 23 33 11.64 | 20 45 03.6 | 17.07 | 0.038 | 15860 | 70 |
| 23 33 40.68 | 20 42 35.6 | 15.48 | 0.036 | 16596 | 42 | 23 35 43.75 | 21 13 13.5 | 17.08 | 0.038 | 41858 | 36 |
| 23 34 57.84 | 21 11 41.8 | 15.51 | 0.036 | 16563 | 28 | 23 33 24.90 | 20 31 53.1 | 17.41 | 0.037 | 19764 | 57 |
| 23 33 10.19 | 20 54 18.8 | 15.56 | 0.037 | 18716 | 102 | 23 33 18.36 | 20 46 58.1 | 17.10 | 0.037 | 16422 | 61 |
| 23 35 45.97 | 21 11 53.9 | 15.59 | 0.037 | 16030 | 40 | 23 32 52.90 | 20 27 37.3 | 17.12 | 0.037 | 42988 | 39 |
| 23 34 23.17 | 20 58 53.0 | 15.62 | 0.036 | 16250 | 35 | 23 34 44.38 | 20 38 13.6 | 17.14 | 0.037 | 16946 | 70 |
| 23 34 35.93 | 20 59 41.0 | 15.63 | 0.037 | 16570 | 36 | 23 32 45.27 | 20 42 13.8 | 17.16 | 0.038 | 19649 | 84 |
| 23 33 32.24 | 20 46 26.9 | 15.63 | 0.037 | 16686 | 47 | 23 34 30.24 | 20 39 38.3 | 17.17 | 0.037 | 33246 | 70 |
| 23 33 57.56 | 21 14 37.7 | 15.67 | 0.037 | 16119 | 40 | 23 33 41.42 | 20 45 41.9 | 17.18 | 0.037 | 15684 | 59 |
| 23 33 15.25 | 20 36 31.0 | 15.68 | 0.036 | 18545 | 70 | 23 33 51.89 | 20 42 05.9 | 17.20 | 0.037 | 15848 | 55 |
| 23 34 01.65 | 20 39 32.2 | 15.76 | 0.037 | 16402 | 70 | 23 33 12.20 | 20 35 20.1 | 17.24 | 0.037 | 19137 | 58 |
| 23 33 58.71 | 21 01 12.4 | 15.76 | 0.036 | 16831 | 39 | 23 35 01.52 | 20 29 02.0 | 17.27 | 0.038 | 27090 | 48 |
| 23 33 55.27 | 20 29 18.2 | 15.83 | 0.037 | 15658 | 48 | 23 32 36.77 | 21 09 24.9 | 17.30 | 0.038 | 15845 | 70 |
| 23 33 56.32 | 21 02 41.5 | 15.83 | 0.036 | 16901 | 83 | 23 32 55.01 | 21 25 35.5 | 17.31 | 0.038 | 29476 | 40 |
| 23 34 22.84 | 20 55 36.7 | 15.85 | 0.037 | 17823 | 55 | 23 33 05.74 | 21 14 21.9 | 17.36 | 0.039 | 11564 | 70 |
| 23 34 08.27 | 20 40 16.9 | 15.85 | 0.037 | 16881 | 84 | 23 32 22.40 | 21 17 08.5 | 17.36 | 0.038 | 16597 | 70 |
| 23 34 39.47 | 21 17 47.0 | 15.87 | 0.037 | 17321 | 45 | 23 33 37.53 | 20 37 38.1 | 17.38 | 0.037 | 19497 | 60 |
| 23 33 47.60 | 20 47 27.0 | 15.90 | 0.037 | 15970 | 70 | 23 34 18.27 | 20 31 00.5 | 17.38 | 0.038 | 19238 | 48 |
| 23 33 26.17 | 20 54 29.4 | 15.91 | 0.037 | 19469 | 35 | 23 33 42.86 | 21 11 20.9 | 17.38 | 0.038 | 16190 | 69 |
| 23 33 56.24 | 20 48 33.0 | 15.93 | 0.037 | 16249 | 40 | 23 35 57.68 | 21 04 08.8 | 17.38 | 0.102 | 21438 | 43 |
| 23 33 30.09 | 20 29 33.6 | 15.95 | 0.037 | 19463 | 44 | 23 32 40.65 | 20 33 04.2 | 17.40 | 0.038 | 43099 | 51 |
| 23 34 51.77 | 21 13 41.4 | 15.97 | 0.037 | 21140 | 43 | 23 33 28.90 | 20 42 04.0 | 17.42 | 0.037 | 16293 | 54 |
| 23 34 43.21 | 20 36 19.6 | 15.99 | 0.037 | 16515 | 40 | 23 33 34.03 | 20 39 49.5 | 17.43 | 0.038 | 48119 | 70 |

| RA (1950) | Decl | $R_{23.5}$ | $\sigma_R$ | $v$ | $\sigma_v$ | RA (1950) | Decl | $R_{23.5}$ | $\sigma_R$ | $v$ | $\sigma_v$ |
|---|---|---|---|---|---|---|---|---|---|---|---|
| 23 33 24.57 | 20 29 09.8 | 15.99 | 0.037 | 19122 | 52 | 23 33 28.39 | 21 17 32.5 | 17.44 | 0.037 | 29057 | 44 |
| 23 32 07.44 | 21 02 15.2 | 16.01 | 0.037 | 16133 | 70 | 23 33 00.39 | 20 30 10.9 | 17.44 | 0.037 | 43317 | 47 |
| 23 33 08.07 | 20 36 43.0 | 16.03 | 0.037 | 14403 | 45 | 23 33 44.64 | 20 34 10.7 | 17.45 | 0.038 | 14739 | 70 |
| 23 35 04.54 | 20 24 07.9 | 16.04 | 0.037 | 17749 | 39 | 23 36 40.80 | 20 49 14.4 | 17.46 | 0.038 | 49618 | 70 |
| 23 33 48.84 | 20 57 56.1 | 16.07 | 0.037 | 16371 | 34 | 23 36 37.31 | 20 46 48.5 | 17.46 | 0.037 | 49399 | 51 |
| 23 35 16.17 | 21 11 20.0 | 16.07 | 0.037 | 21511 | 43 | 23 32 53.18 | 20 35 09.1 | 17.47 | 0.038 | 42445 | 58 |
| 23 32 29.60 | 21 07 08.4 | 16.08 | 0.037 | 19560 | 70 | 23 33 08.04 | 20 44 47.4 | 17.49 | 0.038 | 19087 | 57 |
| 23 33 33.61 | 20 49 36.5 | 16.09 | 0.037 | 19923 | 40 | 23 36 09.63 | 20 55 54.2 | 17.49 | 0.038 | 24929 | 39 |
| 23 33 40.66 | 20 52 18.0 | 16.09 | 0.037 | 18179 | 33 | 23 33 37.76 | 20 32 19.2 | 17.51 | 0.038 | 26175 | 70 |
| 23 34 12.74 | 21 04 26.7 | 16.11 | 0.037 | 16771 | 32 | 23 33 22.14 | 21 22 09.5 | 17.55 | 0.039 | 84935 | 43 |
| 23 35 17.57 | 20 23 35.9 | 16.13 | 0.037 | 16827 | 70 | 23 32 35.52 | 20 30 52.3 | 17.56 | 0.038 | 31079 | 58 |
| 23 34 07.46 | 20 55 03.8 | 16.13 | 0.037 | 17093 | 42 | 23 34 19.38 | 21 21 02.2 | 17.58 | 0.038 | 53878 | 62 |
| 23 34 11.71 | 21 05 34.9 | 16.13 | 0.037 | 16484 | 52 | 23 34 58.45 | 21 05 49.7 | 17.60 | 0.038 | 55185 | 83 |
| 23 33 57.86 | 20 51 11.8 | 16.14 | 0.037 | 16102 | 37 | 23 35 43.77 | 21 06 55.5 | 17.60 | 0.038 | 55281 | 42 |
| 23 34 10.24 | 20 23 47.9 | 16.14 | 0.037 | 17473 | 41 | 23 32 36.75 | 20 37 04.3 | 17.60 | 0.038 | 42361 | 59 |
| 23 35 28.59 | 20 30 03.5 | 16.16 | 0.037 | 25937 | 41 | 23 33 18.95 | 21 10 20.6 | 17.62 | 0.039 | 46382 | 46 |
| 23 34 03.23 | 20 18 04.8 | 16.18 | 0.037 | 20017 | 70 | 23 35 45.67 | 20 30 16.2 | 17.62 | 0.038 | 42515 | 53 |
| 23 33 21.38 | 21 24 47.1 | 16.20 | 0.037 | 29179 | 49 | 23 35 24.36 | 21 03 14.0 | 17.63 | 0.038 | 64524 | 69 |
| 23 33 05.54 | 20 51 53.6 | 16.21 | 0.037 | 16726 | 39 | 23 34 53.94 | 20 59 25.5 | 17.65 | 0.038 | 21275 | 47 |
| 23 33 16.69 | 20 45 30.5 | 16.23 | 0.037 | 19070 | 38 | 23 34 53.22 | 21 16 33.3 | 17.66 | 0.038 | 42827 | 84 |
| 23 34 14.49 | 21 13 33.4 | 16.23 | 0.037 | 9890 | 70 | 23 35 14.05 | 20 51 42.1 | 17.69 | 0.039 | 55211 | 59 |
| 23 35 20.35 | 20 40 57.5 | 16.27 | 0.036 | 20915 | 45 | 23 35 50.80 | 21 08 21.2 | 17.74 | 0.103 | 16547 | 48 |
| 23 33 17.50 | 20 52 03.1 | 16.29 | 0.037 | 18822 | 38 | 23 32 18.22 | 20 33 20.9 | 17.75 | 0.037 | 43268 | 72 |
| 23 34 32.48 | 20 58 26.2 | 16.32 | 0.037 | 15511 | 43 | 23 33 50.43 | 21 16 36.4 | 17.77 | 0.038 | 48019 | 44 |
| 23 34 03.13 | 20 43 30.6 | 16.32 | 0.037 | 15782 | 42 | 23 34 45.39 | 20 33 53.9 | 17.77 | 0.038 | 48031 | 70 |
| 23 35 15.25 | 21 12 35.6 | 16.35 | 0.037 | 21351 | 48 | 23 33 15.05 | 20 27 52.5 | 17.82 | 0.038 | 41395 | 70 |
| 23 33 59.18 | 20 51 19.4 | 16.38 | 0.037 | 16812 | 101 | 23 34 28.95 | 21 12 25.0 | 17.84 | 0.038 | 44640 | 48 |
| 23 36 09.98 | 21 06 58.8 | 16.38 | 0.100 | 18972 | 48 | 23 32 47.00 | 20 34 43.7 | 17.85 | 0.038 | 41395 | 60 |
| 23 34 11.14 | 20 41 59.1 | 16.41 | 0.037 | 16460 | 36 | 23 34 04.11 | 20 37 15.1 | 17.88 | 0.038 | 41997 | 72 |
| 23 34 21.48 | 21 01 20.2 | 16.42 | 0.037 | 16300 | 36 | 23 35 12.02 | 20 40 04.2 | 17.91 | 0.039 | 50158 | 41 |
| 23 32 47.17 | 20 40 18.5 | 16.44 | 0.037 | 19358 | 42 | | | | | | |
| Abell 2440 | | | | | | | | | | | |
| 22 21 22.36 | -1 50 13.1 | [1]14.92 | 0.030 | 26972 | *25 | 22 21 10.62 | -1 55 15.3 | [2]17.45 | 0.032 | 25908 | 40 |
| 22 21 13.19 | -1 54 13.8 | [2]15.44 | 0.030 | 26823 | *27 | 22 21 6.15 | -1 48 51.2 | [1]17.49 | 0.032 | 26100 | 47 |
| 22 21 30.28 | -1 43 30.0 | [3]15.69 | 0.030 | 26208 | 32 | 22 22 0.67 | -1 53 14.6 | [3]17.58 | 0.240 | 26135 | 40 |
| 22 21 42.79 | -1 45 52.9 | [3]15.98 | 0.030 | 27836 | *24 | 22 21 12.65 | -1 54 41.2 | [2]17.61 | 0.033 | 28914 | 72 |
| 22 21 49.98 | -1 46 11.1 | [3]16.23 | 0.030 | 26463 | 34 | 22 21 31.02 | -1 42 37.2 | [3]17.62 | 0.032 | 26519 | 61 |
| 22 21 11.84 | -1 48 30.2 | [1]16.47 | 0.031 | 25943 | 36 | 22 21 22.46 | -1 54 37.0 | [2]17.63 | 0.032 | 26964 | 63 |
| 22 21 26.10 | -1 42 3.8 | [3]16.75 | 0.031 | 27505 | 42 | 22 21 14.89 | -1 54 30.2 | [2]17.75 | 0.034 | 27420 | 43 |
| 22 21 2.69 | -2 4 52.8 | [2]16.89 | 0.240 | 26740 | 68 | 22 22 2.76 | -1 54 27.4 | [3]17.79 | 0.240 | 27795 | 69 |
| 22 21 9.37 | -1 50 21.0 | [1]16.93 | 0.032 | 28609 | 42 | 22 20 50.08 | -1 44 47.8 | [1]17.90 | 0.033 | 27963 | 45 |
| 22 21 28.99 | -1 48 35.4 | [1]16.94 | 0.030 | 25742 | 33 | 22 21 25.64 | -1 51 55.7 | [1]18.00 | 0.033 | 27648 | 53 |
| 22 21 1.91 | -1 49 37.0 | 17.00 | 0.031 | 23156 | 72 | 22 21 9.19 | -1 55 3.9 | [2]18.07 | 0.037 | 28453 | 61 |
| 22 21 25.37 | -1 42 56.4 | [3]17.06 | 0.030 | 26280 | 35 | 22 20 55.21 | -1 44 57.2 | [1]18.15 | 0.035 | 26288 | 52 |
| 22 21 13.15 | -1 55 37.3 | [2]17.09 | 0.032 | 25382 | 38 | 22 21 20.09 | -1 50 53.6 | [1]18.22 | 0.033 | 28111 | 53 |
| 22 21 2.34 | -2 5 38.0 | [2]17.16 | 0.240 | 25059 | 51 | 22 21 13.99 | -1 54 6.8 | [2]18.36 | 0.034 | 26588 | 55 |
| 22 21 40.89 | -1 46 25.4 | [3]17.30 | 0.031 | 27862 | 39 | 22 21 38.35 | -1 46 8.7 | [3]18.39 | 0.034 | 28333 | 54 |

* velocity is variance weighted average of our measurement and previous measurement
# member of (1) partition G1, (2) partition G2, or (3) partition G3